\documentclass[12pt]{reportj}
\usepackage{astron} 
\usepackage{deluxetablej}
\makeatletter
\def\@to{to}
\makeatother
\usepackage{times}
\usepackage{multicol} 
\usepackage{graphicx}
\usepackage{tocloft}
\usepackage{physics}
\usepackage{xcolor}
\usepackage[colorlinks=true, urlcolor=blue, linkcolor=black]{hyperref}
\usepackage[nameinlink]{cleveref} 
\usepackage[all]{hypcap} 
\usepackage{fancyheadings}
\emergencystretch=\maxdimen
\usepackage{datetime}
\newdateformat{ddmonthyyyy}{03 April 2026}

\renewcommand\thesection{\arabic{section}}
\renewcommand\thesubsection{\thesection.\arabic{subsection}}

\def\ssection#1{\setcounter{subsection}{0} \refstepcounter{section} \section*{\hbox to \hsize{\large\bf \arabic{section}. #1\hfill }}\label{sec} \addcontentsline{toc}{section}{\arabic{section}. #1}}
\def\ssubsection#1{\setcounter{subsubsection}{0} \refstepcounter{subsection}\subsection*{\hbox to \hsize{\normalsize\bfseries\itshape \arabic{section}.\arabic{subsection} #1\hfill}}\label{subsec} \addcontentsline{toc}{subsection}{\arabic{section}.\arabic{subsection} #1}}
\def\ssubsubsection#1{\refstepcounter{subsubsection}\subsection*{\hbox to \hsize{\normalsize\it \arabic{section}.\arabic{subsection}.\arabic{subsubsection} #1\hfill}}\label{subsubsec} \addcontentsline{toc}{subsubsection}{\arabic{section}.\arabic{subsection}.\arabic{subsubsection} #1}}

\def\ssectionstar#1{\section*{\hbox to \hsize{\large\bf #1\hfill}} \addcontentsline{toc}{section}{#1}}
\def\ssubsectionstar#1{\subsection*{\hbox to \hsize{\normalsize\bfseries\itshape #1\hfill}} \addcontentsline{toc}{subsection}{#1}}
\def\ssubsubsectionstar#1{\subsection*{\hbox to \hsize{\normalsize\it  #1\hfill}} \addcontentsline{toc}{subsection}{#1}}

\renewcommand{\cftaftertoctitle}{%
\mbox{}\hfill{\normalfont Page}}
\begin{document}

~\\

\vspace{-2.4cm}
\noindent\includegraphics*[width=0.295\linewidth]{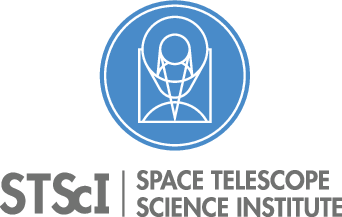}

\vspace{-0.4cm}

\begin{flushright}
    {\bf Instrument Science Report STIS 2026-01(v1)}
    
    \vspace{1.1cm}
    
    {\bf\Huge Verifying the STIS Time Dependent Sensitivity Trends with the Primary CALSPEC Standards}
    
    \rule{0.25\linewidth}{0.5pt}
    
    \vspace{0.5cm}
    
    Daniel Stapleton$^1$, Svea Hernandez$^1$
    \linebreak
    \newline
    \footnotesize{$^1$ Space Telescope Science Institute, Baltimore, MD\\}
    
    \vspace{0.5cm}
    
     \ddmonthyyyy\today 
\end{flushright}

\vspace{0.1cm}

\noindent\rule{\linewidth}{1.0pt}
\noindent{\bf A{\footnotesize BSTRACT}}

{\it \noindent The STIS team monitors the time dependent sensitivity (TDS) of each grating with one from a set of three secondary CALSPEC standard stars: GRW+70D5824, AGK+81D266, and BD+28D4211. 
Here, we use the three primary CALSPEC White Dwarf standard stars, dubbed the standard star ``triad" (GD71, GD153, G191B2B), as an independent set of standards to verify the accuracy of STIS TDS corrections derived from the TDS monitoring stars, 
increasing the sample for each STIS L-mode from one up to three or four standard stars. We focus on triad star observations using the STIS L-mode gratings (e.g., G140L, G230L, etc.) with the same configuration as our standard TDS monitoring programs, and compare the triad observations to the TDS pipeline trends.
Our analysis indicates the relative net count rates inferred from the triad standards agree with the TDS trends derived from the TDS monitoring stars with average residuals $<$ 2$\%$ across the full wavelength range of STIS, suggesting our current TDS L-mode trends are reliable and robust.
We note that the dispersion in the residuals does vary with wavelength, with the NUV showing the lowest spread ($\pm$ 0.32$\%$ at 2400-2500 $\mathrm{\AA}$) and the NIR the largest ($\pm$ 1.32$\%$ at 9500-9900 $\mathrm{\AA}$); however, this scatter is also seen in our measurements of the TDS monitoring stars and 
is more indicative of other instrumental effects. Our findings rule out long term deviations, such as variability in our TDS monitoring stars, within measurement uncertainties.}

\vspace{-0.1cm}
\noindent\rule{\linewidth}{1.0pt}

\renewcommand{\cftaftertoctitle}{\thispagestyle{fancy}}
\tableofcontents


\vspace{-0.3cm}
\ssection{Introduction}\label{sec:Introduction}

The STIS instrument has three operating detectors: the FUV-MAMA, NUV-MAMA, and CCD. Each detector has a set of gratings which cover the broad wavelength range of STIS from 1150 to 10,000 $\mathrm{\AA}$. These gratings
are differentiated by their resolution -- low (L), medium (M), and high (H) -- and wavelength coverage. Time dependent sensitivity (TDS) monitoring is a measure of the changing detector sensitivity since instrument launch. 
TDS corrections are estimated by monitoring the empirical throughput of a particular grating accounting for other sensitivity variations, such as temperature and charge transfer 
inefficiency (CTI). The STIS team monitors the TDS using a selection of representative gratings and a set of standard stars: GRW+70D5824, AGK+81D266, and BD+28D4211, 
with observations occurring in several visits each year. The TDS trends are handled by the \texttt{calstis} pipeline through the TDSTAB reference file.

Monitoring of the TDS trends began shortly after the instrument was turned on (STIS ISR 1998-27). The STIS team routinely updates the TDS corrections used in the pipeline based on the observed throughput changes to maintain the promised flux accuracy of $\sim$4-8$\%$ depending on the configuration (STIS Instrument Handbook Section 16.1). 
The TDS is characterized by the throughput increase or decrease relative to the throughput at the launch date of the instrument. Early TDS analyses sought to track the percent change of each mode per year to
determine accurate throughput corrections (STIS ISRs 1998-28 and 2001-01). Later analyses suggested the need for breakpoints in the TDS trends, indicating a change in the observed rates (STIS ISRs 2014-02 and 2017-06). 
A commonly investigated topic is whether the breakpoints and trends are the same between the L-modes and higher order modes. Analysis of the first TDS data led to the conclusion that the higher order modes follow the same 
trends as the L-modes, and so the L-mode corrections were then applied for all other modes in the \texttt{calstis} pipeline (STIS ISR 2001-01). Later analysis comparing the TDS trends observed for the L-modes against those from the higher-order modes confirmed the general agreement, 
with the exception of G430M (STIS ISRs 2014-02 and 2017-06). Given the historical importance of the L-mode trends and that they are applied to all STIS observing modes, an independent verification of the TDS monitoring stars is warranted.

STIS spectrophotometric fluxes provide the observational backbone of the CALSPEC flux standard library (Bohlin et al. 2014).  However, STIS data in CALSPEC is maintained with an independent data reduction pipeline, including independently derived absolute flux calibration 
and TDS trends that rely solely on the observations of the primary standard stars GD71, GD153, and G191B2B (Bohlin, R. C. 2014 and 2018), which we refer to collectively as the ``triad." The triad stars have been observed in a manner similar to our TDS monitoring programs since the launch 
of STIS, but the sensitivity trends derived from these stars have not been directly compared with those applied in \texttt{calstis} via the TDSTABs. Here, we use triad stars to measure the TDS trends using the same analysis tools as for our normal TDS monitoring stars.

\lhead{}
\rhead{}
\cfoot{\rm {\hspace{-1.9cm} Instrument Science Report STIS 2026-01(v1) Page \thepage}}

\begin{deluxetable}{lcccr}
  \tabcolsep 4pt
  \tablewidth{0pt}
  \tablecaption{All of the data used in this analysis organized by corresponding cycle, proposal ID, grating, and the target(s) observed by that grating are listed. All of the observations use aperture 52x2. \label{tab: Table 1}}
  \tabletypesize{\footnotesize}
  \tablehead{
    \colhead{Cycle} &  \colhead{Proposal ID} & \colhead{STIS Grating}  & \colhead{Target(s)}}
  \startdata
  
  7 & 7063  & G230LB &  GD153 \\
               &   & G430L &  GD153 \\
               &   & G750L &  GD153 \\
               \hline \vspace{-0.3cm} \\     
  7 & 7095  & G230LB  &  GD153 \\
               \hline \vspace{-0.3cm} \\
  7 & 7096 & G140L  & GD153 \\
              &   & G230L  & GD153 \\
              \hline \vspace{-0.3cm} \\  
  7 & 7097 & G140L  & GD153 \\
              &   & G230L  & GD153 \\
              \hline \vspace{-0.3cm} \\  
  7 & 7656 & G140L  & GD71, GD153 \\
              &   & G230L  & GD71, GD153 \\
              &   & G230LB  & GD71, GD153 \\ 
              &   & G430L  & GD71, GD153 \\ 
              &   & G750L  & GD71, GD153 \\ 
              \hline \vspace{-0.3cm} \\  
  7 & 7657 & G230LB  & GD71 \\
              \hline \vspace{-0.3cm} \\  
  7 & 7674 & G750L  & GD71, G191B2B \\
              \hline \vspace{-0.3cm} \\  
  7 & 7805 & G230LB  & GD153, G191B2B \\
              &   & G430L  & GD153, G191B2B \\
              &   & G750L  & GD153, G191B2B \\ 
              \hline \vspace{-0.3cm} \\  
  7 & 7917 & G140L  & GD71 \\
              &   & G230L  & GD71 \\
              \hline \vspace{-0.3cm} \\  
  7 & 7932 & G140L  & GD71 \\
              \hline \vspace{-0.3cm} \\  
  7 & 7937 & G140L  & GD71 \\
              &   & G430L  & G191B2B \\
              \hline \vspace{-0.3cm} \\  
  7 & 8016 & G140L  & GD153 \\
              &   & G230L  & GD153 \\
              \hline \vspace{-0.3cm} \\  
  8 & 8421 & G140L  & GD71 \\
              &   & G230L  & GD71 \\
              &   & G230LB  & GD71 \\ 
              &   & G430L  & GD71 \\
              &   & G750L  & GD71 \\
              \hline \vspace{-0.3cm} \\  
  8 & 8505 & G140L  & GD71 \\
              &   & G230L  & GD71 \\
              &   & G230LB  & GD71 \\ 
              &   & G430L  & GD71 \\ 
              &   & G750L  & GD71 \\ 
              \hline \vspace{-0.3cm} \\  
  9 & 8849 & G230LB  & G191B2B \\
              &   & G430L  & G191B2B \\
              &   & G750L  & G191B2B \\ 
              \hline \vspace{-0.3cm} \\  
  10 & 8915 & G230LB  & G191B2B \\
              \hline \vspace{-0.3cm} \\  
  10 & 8916 & G140L  & GD71, GD153 \\
              &   & G230L  & GD71, GD153 \\
              &   & G230LB  & GD71, G191B2B \\ 
              &   & G430L  & GD71 \\ 
              &   & G750L  & GD71, G191B2B \\ 
              \hline \vspace{-0.3cm} \\  
  12 & 10039 & G140L  & GD71, GD153 \\
              &   & G230L  & GD71, GD153 \\
              &   & G230LB  & GD71, GD153, G191B2B \\ 
              &   & G430L  & GD71, GD153, G191B2B \\ 
              &   & G750L  & GD71, GD153, G191B2B \\ 
              \hline \vspace{-0.3cm} \\  
  17 & 11393 & G140L  & GD153 \\
              \hline \vspace{-0.3cm} \\
  17 & 11394 & G230L  & GD153 \\
              \hline \vspace{-0.3cm} \\
  17 & 11403 & G140L  & GD153 \\
              &   & G230L  & GD153 \\
              \hline \vspace{-0.3cm} \\
  17 & 11889 & G230LB  & G191B2B \\
              &   & G430L  & G191B2B \\
              &   & G750L  & G191B2B \\ 
              \hline \vspace{-0.3cm} \\
  17 & 11999 & G140L  & GD71, GD153 \\
              &   & G230L  & GD71, GD153 \\
              &   & G230LB  & GD71, GD153 \\
              &   & G430L  & GD71, GD153 \\
              &   & G750L  & GD71, GD153 \\ 
              \hline \vspace{-0.3cm} \\
  18 & 12392 & G230LB  & G191B2B \\
              &   & G430L  & G191B2B \\
              &   & G750L  & G191B2B \\ 
              \hline \vspace{-0.3cm} \\
  19 & 12682 & G140L  & GD153 \\
              &   & G230L  & GD153 \\
              &   & G230LB  & GD153 \\
              &   & G430L  & GD153 \\
              &   & G750L  & GD153 \\ 
              \hline \vspace{-0.3cm} \\
  19 & 12737 & G140L  & GD71 \\
              &   & G230L  & GD71 \\
              &   & G230LB  & GD71, G191B2B \\ 
              &   & G430L  & GD71, G191B2B \\
              &   & G750L  & GD71, G191B2B \\
              \hline \vspace{-0.3cm} \\
  19 & 12813 & G230LB  & GD71, G191B2B \\
              &   & G430L  & GD71, G191B2B \\
              &   & G750L  & GD71, G191B2B \\ 
              \hline \vspace{-0.3cm} \\
  20 & 13162 & G140L  & GD153 \\
              &   & G230L  & GD153 \\
              &   & G230LB  & GD153 \\ 
              &   & G430L  & GD153 \\ 
              &   & G750L  & GD153 \\ 
              \hline \vspace{-0.3cm} \\
  21 & 13599 & G140L  & GD71, GD153 \\
              &   & G230L  & GD71, GD153 \\
              &   & G230LB  & GD71, GD153, G191B2B \\
              &   & G430L  & GD71, GD153, G191B2B \\ 
              &   & G750L  & GD71, GD153, G191B2B \\  
              \hline \vspace{-0.3cm} \\
  24 & 14861 & G140L  & GD71, GD153 \\
              &   & G230L  & GD71, GD153 \\
              &   & G230LB  & GD71, GD153, G191B2B \\ 
              &   & G430L  & GD71, GD153, G191B2B \\
              &   & G750L  & GD71, GD153, G191B2B \\
              \hline \vspace{-0.3cm} \\
  26 & 15602 & G140L  & GD71, GD153 \\
              &   & G230L  & GD71, GD153 \\
              &   & G230LB  & GD71, GD153, G191B2B \\ 
              &   & G430L  & GD71, GD153, G191B2B \\
              &   & G750L  & GD71, GD153, G191B2B \\
              \hline \vspace{-0.3cm} \\
  28 & 16436 & G140L  & GD71, GD153 \\
              &   & G230L  & GD71, GD153 \\
              &   & G230LB  & GD71, GD153, G191B2B \\ 
              &   & G430L  & GD71, GD153, G191B2B \\
              &   & G750L  & GD71, GD153, G191B2B \\
              \hline \vspace{-0.3cm} \\
  28 & 16438 & G140L  & G191B2B \\
              \hline \vspace{-0.3cm} \\
  \enddata
\end{deluxetable}

\vspace{-0.3cm}

\ssection{Observations}\label{sec:Observations}

The triad is observed every other cycle, one exposure per triad star per L-mode, by PI Ralph Bohlin since 2016 (e.g., HST program ID 15602, 16436, 16966). 
Prior to these programs, the triad had been observed every few years dating back to the launch of STIS in 1997 (e.g., HST program ID 7656). These programs observed the triad with 
the same configuration as our current TDS monitoring programs, i.e., using the same apertures, gratings, and cenwaves. However, the suite of observations of the triad is much smaller than the suite of observations of the TDS monitoring stars.
We prioritized selecting as much data as possible for each star and grating that satisfied predefined criteria: the desired star, grating with a central wavelength corresponding to the nominal value for that grating, and an 
aperture of 52x2. The 52x2 aperture is used because it has the highest photometric accuracy (STIS Instrument Handbook Section 13.4). We excluded exposures with zero second exposure times and organized data sets into 
those with exposures $>$ 200 seconds and those $<$ 200 seconds. This was done to study the effect of lower signal-to-noise (S/N) observations on the uncertainties of the trends. Our TDS analysis relied on the pipeline
\texttt{x1d.fits} and \texttt{sx1.fits} files; however, we also included the \texttt{flt.fits} files for examinations of cosmic ray rejection as well as the corresponding \texttt{sx2.fits}
and \texttt{raw.fits} for completeness. In \Cref{tab: Table 1}, we list the proposal information for the data analyzed here. The proposals listed also include data taken with the 52x2E1 aperture, which we exclude from this analysis.

\begin{figure}[!h]
  \centering
  \includegraphics[width=\textwidth]{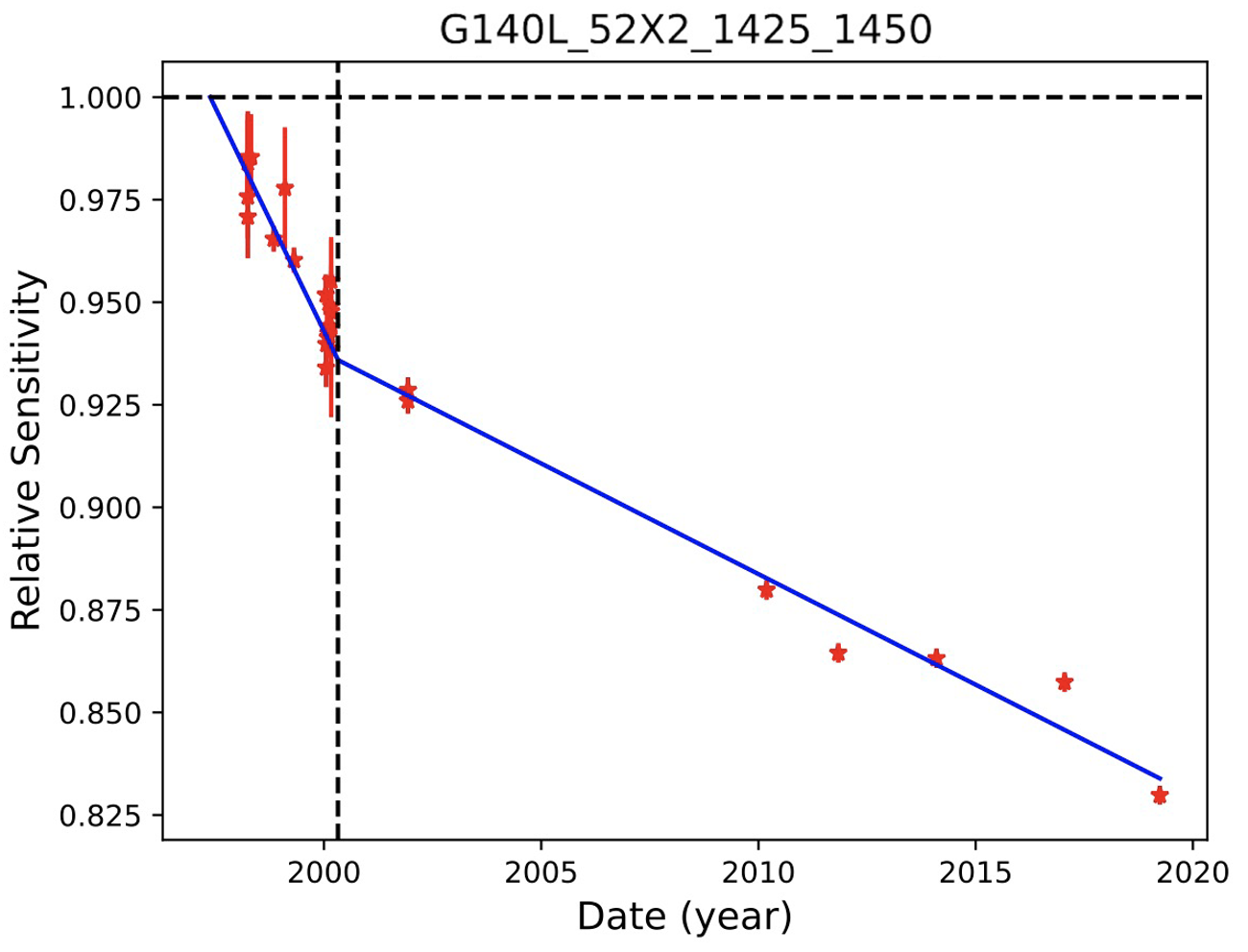}
    \caption{The relative sensitivity as a function of time for G140L data of GD71. The solid blue line indicates a test fit performed by the code, but not used in our analysis. The vertical dotted black line indicates a break point in the fit. All error bars are included. The title indicates the: grating, aperture, central wavelength of grating, and starting wavelength of the bin in $\mathrm{\AA}$.}
    \label{fig: Figure 1}
\end{figure}

\begin{figure}[!h]
  \centering
  \includegraphics[width=\textwidth]{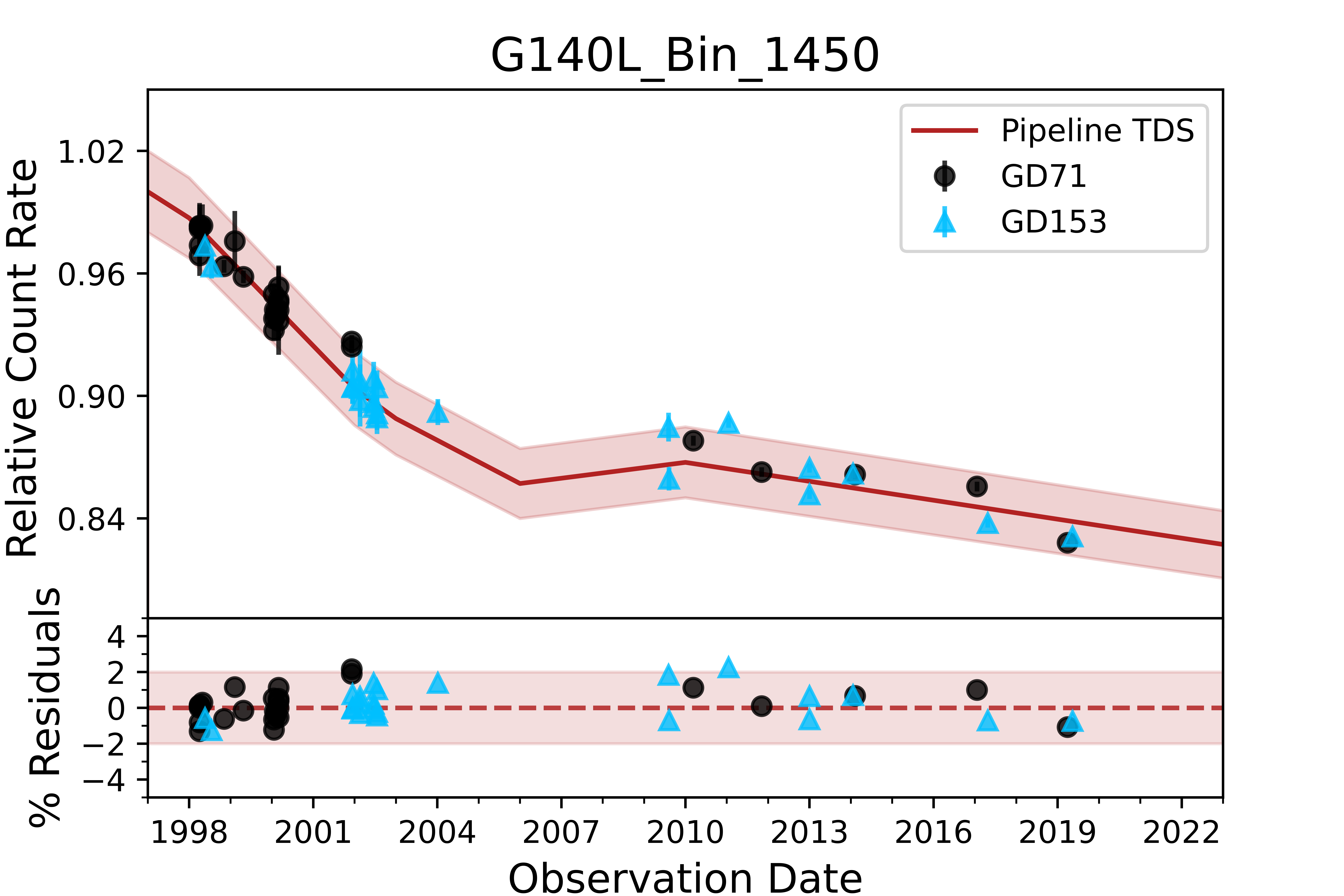}
    \caption{A time dependent sensitivity plot using G140L data of the triad stars GD71 and GD153, for the bin of starting wavelength 1450 $\mathrm{\AA}$. All exposures use aperture 52x2. The red trend line indicates the adopted pipeline TDS from our TDS monitoring standard stars. The shaded region indicates $\pm$2$\%$ around the adopted TDS trend. Error bars are included.}
    \label{fig: Figure 2}
\end{figure}

\vspace{-0.3cm}
\ssection{Methods}\label{sec:methods}
The standard TDS analysis software performs the following steps: 
\begin{enumerate}
  \item The net count rates for every observation are corrected in order to isolate the effects of the TDS. Those corrections include charge transfer inefficiency (CTI) for the CCD, red halo correction for G750L, and temperature-dependent sensitivity for the CCD and FUV-MAMA.  
  \item The corrected net count rates and uncertainties for each exposure are separated into wavelength bins spanning $\pm$ 50, 100, or 200 $\mathrm{\AA}$ depending on the grating (see Table 4 of STIS ISR 2014-02 for binning information). The TDS of each star and low-order grating will therefore be characterized by 11--15 different wavelength bins. 
  \item A weighted sum of the net count rates in each bin is calculated, where the weight is determined by the number of pixels in a bin. The uncertainties are then propagated from the weighted sum and calculated using the \texttt{ERROR} array in the \texttt{x1d.fits} and \texttt{sx1.fits} files.
  \item The summed net count rates and uncertainties for each bin are then plotted as a function of time for exposures with matching star, grating, and aperture. 
  \item A linear fit is performed to estimate the change rate ($\%$/yr). The fit relies on the adopted reference time (year 1997.38 = MJD 50587.0) as a fitting parameter and iterates using a segmented line model (see STIS ISR 2017-06 for a brief description of the fitting procedure and STIS ISR 2014-02 for a more in-depth description of how the time breakpoints are derived). 
        Based on the fit, the net count rates are all normalized by the reference time fitting parameter, such that the TDS at the reference time = 1. An example plot of the fitted sensitivity data is shown in \Cref{fig: Figure 1}.  
  \item Finally, the normalized TDS net count rates and uncertainties are plotted alongside the adopted \texttt{calstis} pipeline trend for comparison. An example of this step is shown in \Cref{fig: Figure 2}.
\end{enumerate}

\vspace{-0.3cm}
\ssection{Data Reduction}\label{sec:data_reduction}
Our sample sizes for each triad star and grating are relatively small with $\sim$20 observations compared to hundreds for our TDS monitoring stars. However, not all of the exposures were taken with nominal configurations. Some of them have slightly
different parameters, such as different MSM offsets (for the MAMAs) or detector gains (for CCD observations). Additionally, several exposures required recalibration (e.g., to improve cosmic ray rejection). We corrected these issues where possible, 
and in this section we describe the improved reduction. This discussion follows the order in which the gratings were analyzed, from shortest to longest wavelength. 
\ssubsection{FUV and NUV-MAMA Reduction}\label{sec:mama_reduction}
\ssubsubsection{POSTARG Issues}\label{sec:postarg_issues}
Starting with G140L, we identified a few observations not centered at the nominal position on the detector. Offsets from the nominal position are recorded
in the header keywords as \texttt{POSTARG1} and \texttt{POSTARG2}, corresponding to shifts in the dispersion and cross-dispersion directions. A nominal exposure has \texttt{POSTARG1} = 0.0 and \texttt{POSTARG2} = 0.0. For consistency with the standard TDS analysis, we include only datasets at the nominal 
position in the cross-dispersion direction at \texttt{POSTARG2} = 0.0, as a non-zero \texttt{POSTARG2} exposure would place the target spectrum on a different part of the detector, which might require a different TDS characterization.
\ssubsubsection{MSM Offset Issues}\label{sec:msm_issues}
An alternative way of moving the location of the spectrum on the detector is by tilting the gratings with an offset of the mode selection mechanism (MSM).  The target is still at the center of the aperture, but the image of the aperture is shifted on the detector.  
Monthly offsets of the MSM are applied to all G140L and G230L GO data to more evenly deplete charge on the MAMAs, but special commanding to disable this offsetting is routinely applied to TDS datasets. We found that triad star data taken for Cycle 7 HST Program 8016 contained non-zero MSM offsets, and these were excluded from the analysis.
\begin{figure}[!h]
  \centering
  \includegraphics[width=\textwidth]{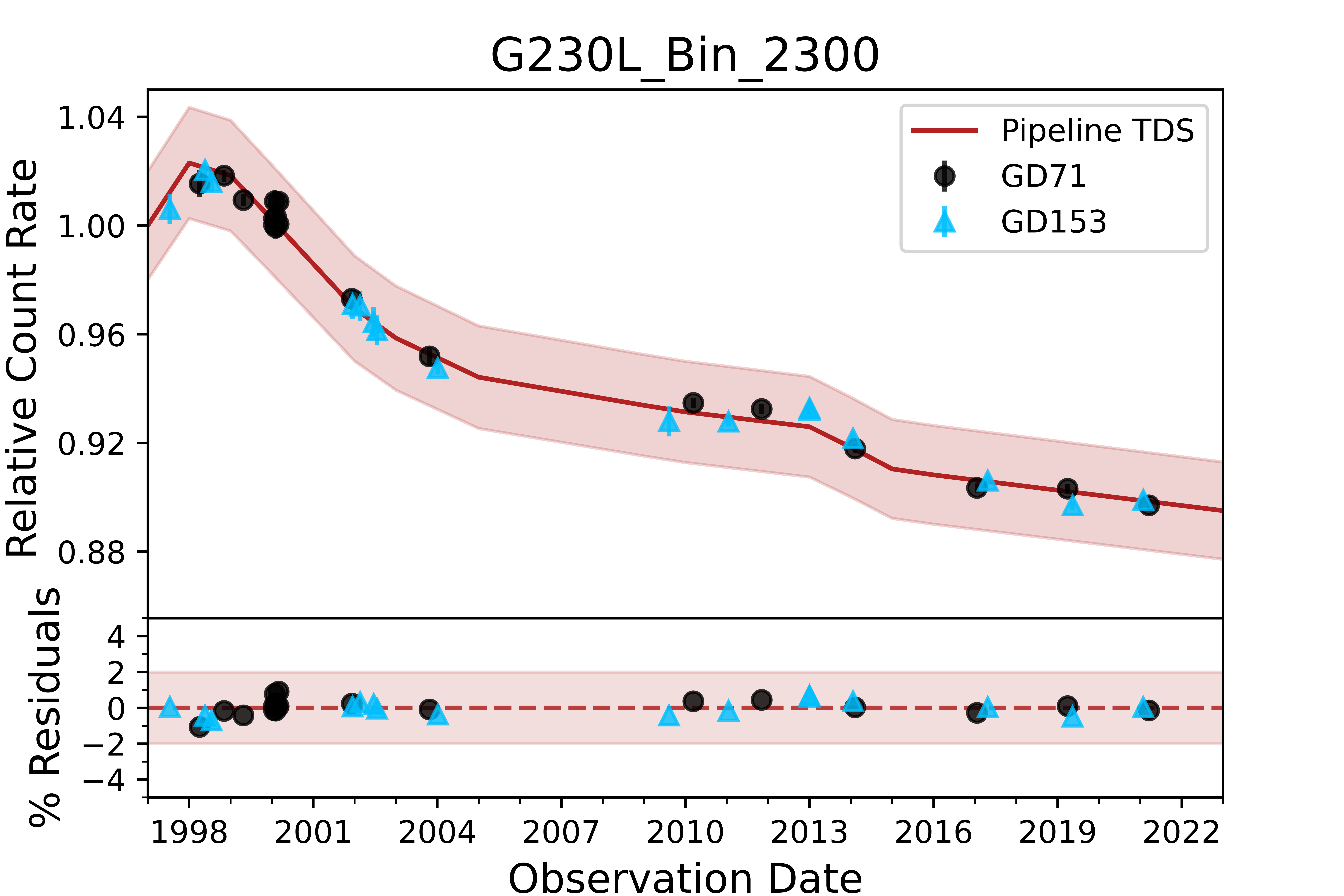}
    \caption{The same caption as \Cref{fig: Figure 2}, but for the G230L bin of starting wavelength 2300 $\mathrm{\AA}$.}
    \label{fig: Figure 3}
\end{figure}
\ssubsection{CCD Reduction}\label{sec:ccd_reduction}
The following sub-sections pertain to the CCD gratings (G230LB, G430L, and G750L):
\ssubsubsection{Gain Changes}\label{sec:gain_changes}
Unlike the primary TDS monitoring observations where all are taken with a gain of 1 (\texttt{CCDGAIN} = 1), the exposures in this suite of observations have a mix of \texttt{CCDGAIN} = 1 and \texttt{CCDGAIN} = 4 exposures. A direct comparison of the raw net count rates between the two types of datasets 
would therefore be off by a factor of $\sim$4. To resolve this issue, we multiplied the net count rates of the \texttt{CCDGAIN} = 4 exposures by the actual gain of the CCD (\texttt{ATODGAIN}) found in the header of each exposure.
\begin{figure}[!h]
  \centering
  \includegraphics[width=\textwidth]{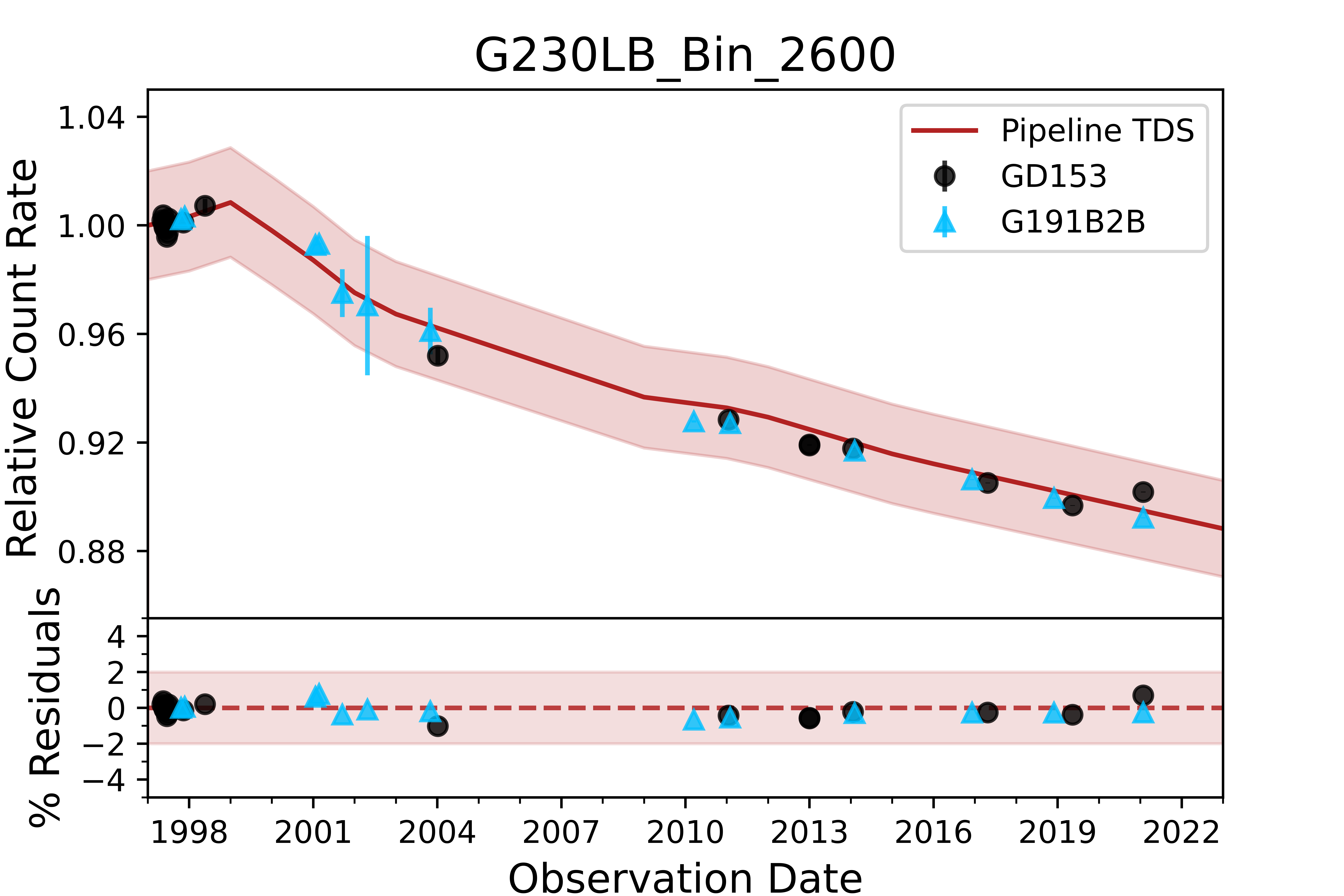}
    \caption{The same caption as \Cref{fig: Figure 2}, but for the G230LB bin of starting wavelength 2600 $\mathrm{\AA}$, and triad stars GD153 and G191B2B.}
    \label{fig: Figure 4}
\end{figure}
\ssubsubsection{Cosmic Ray (CR) Overcorrection}\label{sec:cr_overcorrect}
A recurring issue with each CCD grating was individual exposures having suspiciously low relative net count rates. In STIS data, a cause of this could be an overcorrection for cosmic rays reducing the overall net count rate in an exposure. This issue was investigated in STIS ISR 2019-02, 
where it was found that in the cosmic ray rejection process of the pipeline, datasets with misaligned \texttt{CR-SPLIT} sub-exposures removed pixels preferentially in the spectral extraction region. This is because the peaks of the line spread functions are slightly shifted between the sub-exposures. In \texttt{CR-SPLIT} = 2 datasets 
in particular, the count difference due to the spatial shift causes the pixels with higher (real) counts to instead be flagged as cosmic rays and removed from the combined image.

We checked for this phenomenon by calculating the percentage of pixels removed from the extraction region as compared with the background, with greater than 2--3$\%$ more pixels removed in the extraction region noticeably reducing the net count rate and therefore the flux.  We inspected individual
exposures in our TDS trends using \texttt{crrej\_exam.py} (STIS ISR 2019-02) and found that in a few instances the CR pixel removal in the extraction region was as high as 30$\%$. We ran these exposures through a CR re-correction code -- the \texttt{stistools} module \texttt{ocrreject.py} -- where we changed the \texttt{CRSIGMAS} of CR 
selection and set the \texttt{INITGUES} to `min' as opposed to `median' in the \texttt{calstis} pipeline.\footnotemark{} We tweaked \texttt{CRSIGMAS} to 10 or higher until the CR pixel rejection in the extraction region was within 2$\%$ of the rest of the detector. The maximum \texttt{CRSIGMAS} needed was 22.
The data reduced with the updated CR rejection saw net count rates increase by as much as 8$\%$ and typically brought them in line with the TDS trends of the unaffected exposures.

\footnotetext{The \texttt{stistools} documentation can be found at: \url{https://stistools.readthedocs.io/en/latest/index.html}. Note that \texttt{ocrreject\_exam.py} has replaced \texttt{crrej\_exam.py}.}

\begin{figure}[!h]
  \centering
  \includegraphics[width=\textwidth]{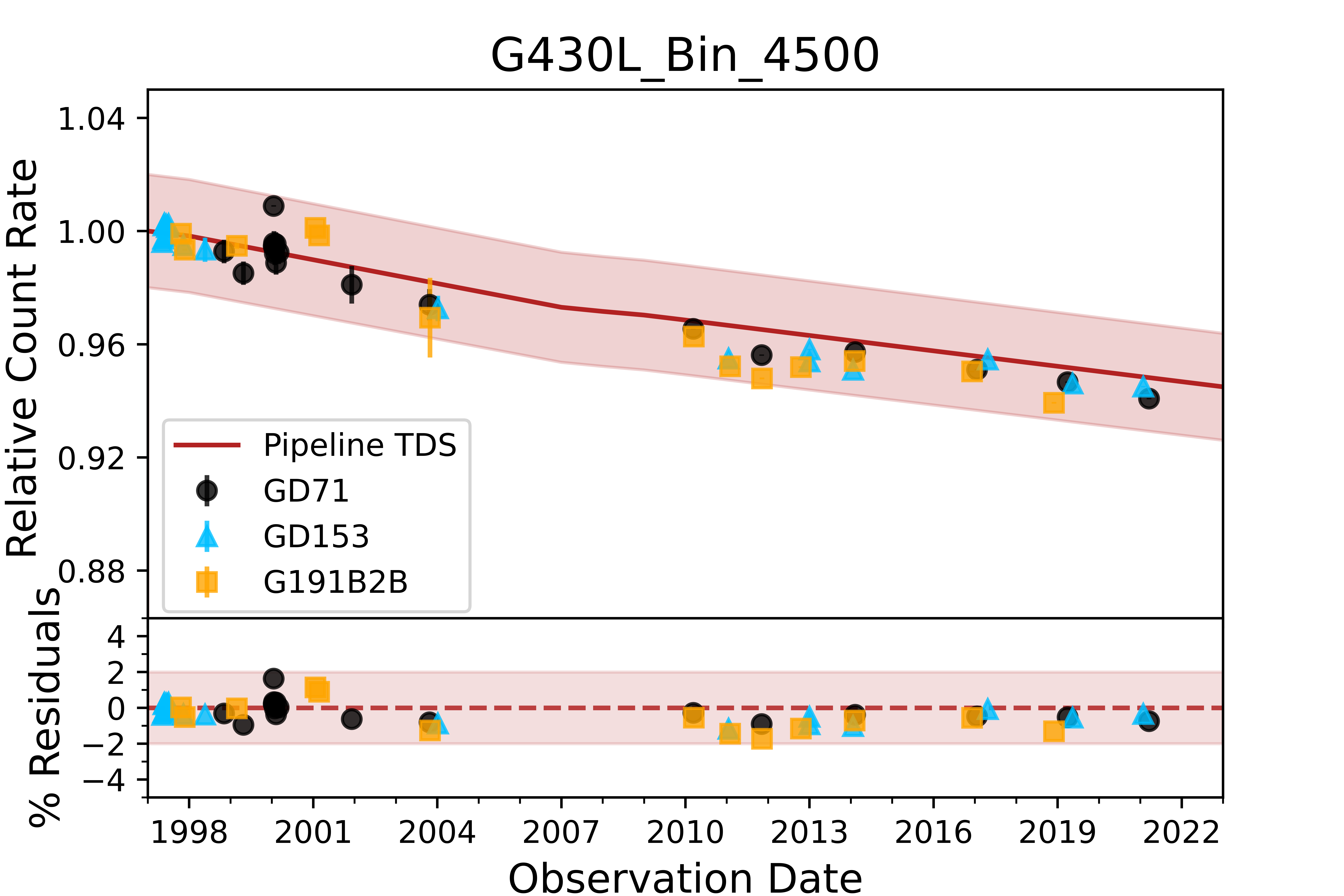}
    \caption{The same caption as \Cref{fig: Figure 2}, but for the G430L bin of starting wavelength 4500 $\mathrm{\AA}$, and all three triad stars.}
    \label{fig: Figure 5}
\end{figure}

\begin{figure}[!h]
  \centering
  \includegraphics[width=\textwidth]{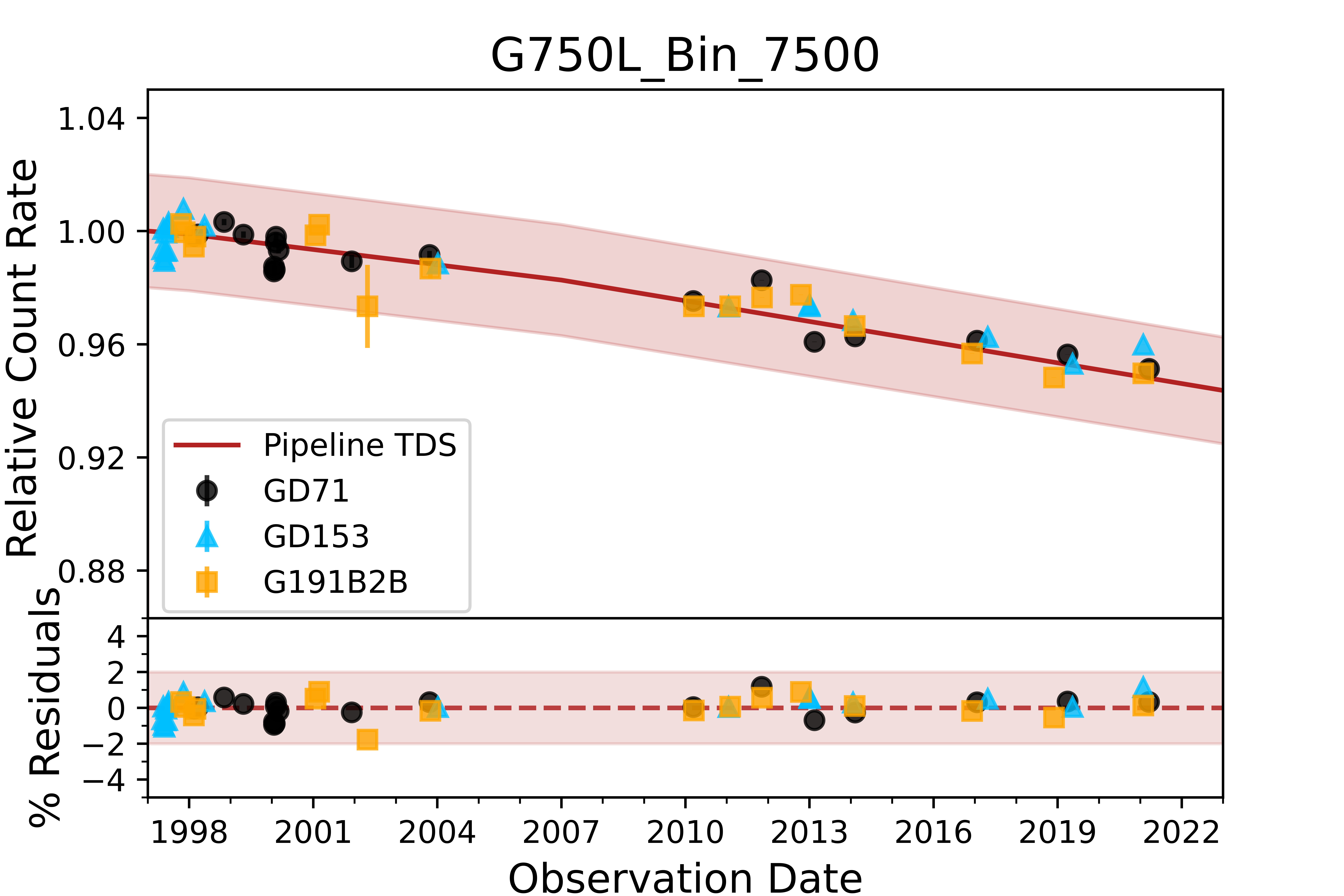}
    \caption{The same caption as \Cref{fig: Figure 2}, but for the G750L bin of starting wavelength 7500 $\mathrm{\AA}$, and all three triad stars.}
    \label{fig: Figure 6}
\end{figure}

\vspace{-0.3cm}
\ssection{Results and Conclusions}\label{sec:results_and_conclusion}
Figures 2--6 show representative plots of our final TDS comparisons for each of the 5 L-mode gratings. They show the measured TDS trends of the triad stars compared to the pipeline TDS trends derived from our TDS monitoring stars. Appendix A has an expanded set of figures for each grating that includes three bins: the shortest and longest available wavelength bins and one intermediate wavelength bin. 
Our conclusions are based on residuals between the relative count rate of each triad star and the TDS pipeline corrections (residual = (relative count rate - pipeline) / pipeline), as can be seen in the “$\%$ Residuals” subplot of our TDS figures.  We compare the average of the residuals for a given wavelength bin, such as the 2600 $\mathrm{\AA}$ bin of G230LB in \Cref{fig: Figure 4}, in order to determine how close the residual values are to the pipeline trends.
An average residual of 0$\%$ indicates that the relative count rates for that wavelength bin exactly matches the pipeline fit, and an average residual outside $\pm$2$\%$ would indicate that the relative count rates are outside the STIS team's threshold for an accurate TDS pipeline correction. We also calculated the standard deviation of the residuals to quantify the spread of the triad observations about the TDS pipeline corrections.\footnotemark{}
The average residuals and standard deviations for each wavelength bin have been tabulated in \Cref{tab: Table 2} in Appendix B. 

\footnotetext{For a more robust statistical analysis, we can use these standard deviations, and rely on the assumption that STIS data follow a normal distribution in the high count regime ($\sim$$>$10 counts) (STIS Data Handbook Section 5.5.4). Under this treatment, the standard deviation ($\sigma$) of a mean residual being 2$\sigma$ away from 2$\%$, would indicate an $\sim$95$\%$ probability that the sensitivity trend of the count rates in
a given wavelength bin is within our $\pm$2$\%$ threshold. Relying on \Cref{tab: Table 2}, we can see that this is true for the majority of data samples (e.g., G230LB observations of GD153), and wavelength bins which are $<$2$\sigma$ from having a $\geq$2$\%$ average residual follow expected behavior as described in this section.}

The average residuals of the G140L bins are consistently between $\sim$0.10--0.30$\%$, but the shortest and longest wavelength bins have standard deviations on the order of 1$\%$ (e.g., 0.09 $\pm$ 1.11$\%$ for the 1150 $\mathrm{\AA}$ bin).
The average residuals of the G230L bins have similarly large standard deviations in the shortest wavelength bin, such as the 1600 $\mathrm{\AA}$ bin with an average residual of -0.26 $\pm$ 1.31$\%$ (see \Cref{fig: Figure 8} in Appendix A), but otherwise have average residuals at $\sim$0.5$\%$. G230LB behaves similarly to G230L with the highest deviation occurring in the 1700 $\mathrm{\AA}$ bin (-0.02 $\pm$ 0.96$\%$), and the majority of bins constrained to within 0.5$\%$.
The G430L residuals are also consistently constrained as compared with our $\pm$2$\%$ threshold at about 0$\%$; however, there is a consistent negative offset relative to the trend (see \Cref{fig: Figure 10} in Appendix A), which requires further investigation. Finally, G750L shows excellent agreement (for example, 0.06 $\pm$ 0.5$\%$ for the 5900 $\mathrm{\AA}$ bin) until reaching the reddest wavelengths at 7900 $\mathrm{\AA}$ and onward, where the standard deviation increases to $>$1$\%$, 
such as the 8700 $\mathrm{\AA}$ bin with an average residual of 0.39 $\pm$ 1.2$\%$. 

In all cases, our TDS trends derived from the triad of standard stars either match the TDS calibrations adopted in the pipeline to within the $\pm$2$\%$, or, where the standard deviations are largest, exhibit the same behavior as our TDS monitoring stars. 
For example, a few bins of G140L (FUV-MAMA) and G230L (NUV-MAMA) have standard deviations $>$1$\%$ (e.g., $\pm$1.26$\%$ in the 1550--1600 $\mathrm{\AA}$ bin of G140L), indicating a larger spread of the relative count rates around the pipeline TDS trends, whereas most bins of the CCD have standard deviations $<$1$\%$. This behavior is similarly observed in our TDS monitoring programs and is, therefore, expected.\footnotemark{} In G750L, an increasing relative sensitivity at the reddest wavelengths (see \Cref{fig: Figure 11}c in Appendix A) is
the result of a lack of fringe correcting in our TDS code (STIS ISR 2017-06). This behavior matches what we see from the TDS monitoring stars and again encourages further investigation. The pipeline trends may require updates in those regimes based on observations of the TDS monitoring stars.
We note that in some cases the limited number of exposures taken with nominal settings makes this type of analysis more challenging. In particular, observations of GD71 taken with G230LB were taken predominantly pre-Servicing Mission 4 (pre-SM4) and do not provide
a complete time coverage of the TDS. We highlight, however, that the available exposures are qualitatively similar to the adopted TDS pipeline trends. There are also 52x2E1 aperture observations available which align well with the pipeline trend in the post-SM4 era, suggesting G230LB observations of GD71 show overall agreement with the pipeline, as well as supporting a scenario where the sensitivity at the nominal and E1 positions are comparable to each other.

\footnotetext{The latest TDS plots from our TDS monitoring programs may be found at: \url{https://www.stsci.edu/hst/instrumentation/stis/performance/sensitivity}.}

Our analysis indicates the TDS trends adopted from observations of GRW+70D5824 and AGK+81D266 robustly describe the TDS of the STIS L-modes. This analysis also rules out long term deviations, such as variability in our TDS monitoring stars, within measurement uncertainties. The team, therefore, will continue using GRW+70D5824 and AGK+81D266 to monitor the L-modes as they have provided an accurate characterization of the TDS. The STIS team historically applies the L-mode 
analysis to the M-modes and echelles, but as mentioned in STIS ISR 2017-06, uncertainties in the accuracy of this method will be investigated (e.g., deviations from the \texttt{calstis} pipeline trends described in STIS ISR 2004-04). The team will continue monitoring the three secondary standard stars in this report for independent validation. 


\vspace{-0.3cm}
\ssectionstar{Acknowledgements}
\vspace{-0.3cm}
Special thanks to TalaWanda Monroe for her detailed review of this report, and Sean Lockwood for his investigation into the process for hot pixel removal.

\vspace{-0.3cm}
\ssectionstar{Change History for STIS ISR 2026-01}\label{sec:History}
\vspace{-0.3cm}
Version 1: \ddmonthyyyy{03 April 2026} - Original Document 

\vspace{-0.3cm}
\ssectionstar{References}\label{sec:References}
\vspace{-0.3cm}

\noindent
Bohlin, R. C. 2014 $AJ$ \textbf{147} 127
\\
Bohlin, R. C. 2018, \textit{Proceedings of the International Astronomical Union}, \textbf{14} (A30) 449-453
\\
Bohlin, R. C., $\&$ Proffit, C. R. 2015, STIS ISR 2015-01
\\
Carlberg, J., $\&$ Monroe, T. 2017, STIS ISR 2017-06
\\
Carlberg, J. 2019, STIS ISR 2019-02
\\
Downes, R., $\&$ Hodge, P. 1998, STIS ISR 1998-28
\\
Hernandez, S. 2021, STIS ISR 2021-01
\\
Holland et al. 2014, STIS ISR 2014-02
\\
Landolt, A.U. 1992 $AJ$ \textbf{104} 1
\\
Landolt, A.U. 2013 $AJ$ \textbf{146} 131
\\
Leitherer, C. 1997, ``Effect of MAMA Charge Offsetting on Sensitivity and Dispersion Accuracy," HST Proposal ID 7917, Cycle 7
\\
Prichard, L., Welty, D. and Jones, A., et al. 2022 ``STIS Instrument Handbook," Version 21.0, (Baltimore: STScI)
\\
Rickman, E., Brown J., et al. 2024, ``STIS Data Handbook", Version 8.0, (Baltimore: STScI)
\\
Stys, D., $\&$ Walborn, N. 2001, STIS ISR 2001-01
\\
Walborn, N., $\&$ Bohlin, R. 1998, STIS ISR 1998-27

\vspace{-0.3cm}
\ssectionstar{Appendix A}\label{sec:Appendix_A}
\vspace{-0.3cm}
All the following figures give three example TDS comparison plots for each available star and each grating. For comparison, we chose the same three wavelength bins for stars observed with the same grating. 

\begin{figure}[!htbp]
  \centering
  \includegraphics[width=\textwidth]{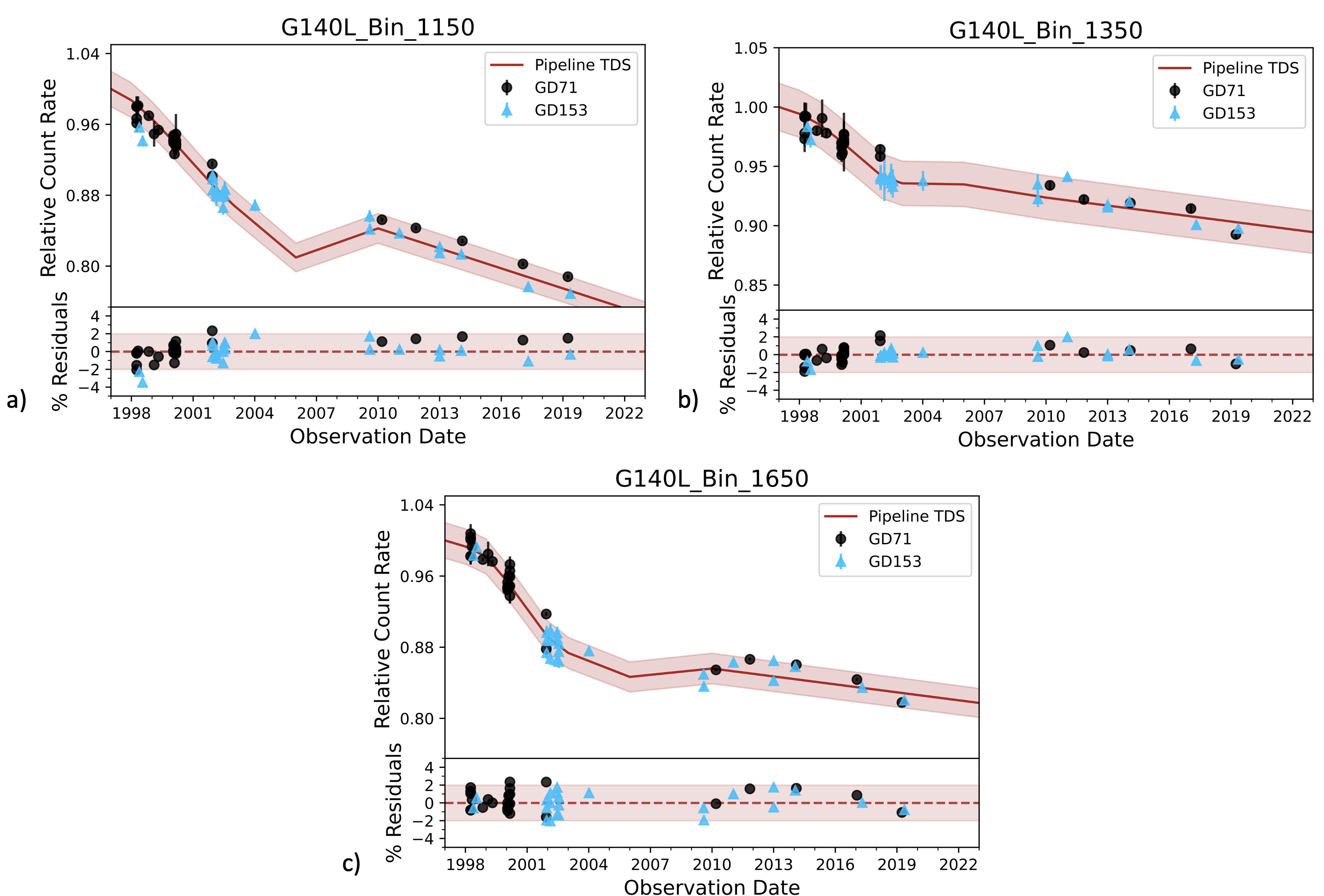}
    \caption{These are three time dependent sensitivity plots using G140L data of the triad stars GD71 and GD153. Each panel corresponds with a bin of starting wavelength: a) 1150 $\mathrm{\AA}$, b) 1350 $\mathrm{\AA}$, and c) 1650 $\mathrm{\AA}$, respectively. All exposures use aperture 52x2. The red trend line indicates the adopted pipeline TDS from our TDS monitoring standard stars. The shaded region indicates $\pm$2$\%$ around the adopted TDS trend. Error bars are included.}
    \label{fig: Figure 7}
\end{figure}

\begin{figure}[!htbp]
  \centering
  \includegraphics[width=\textwidth]{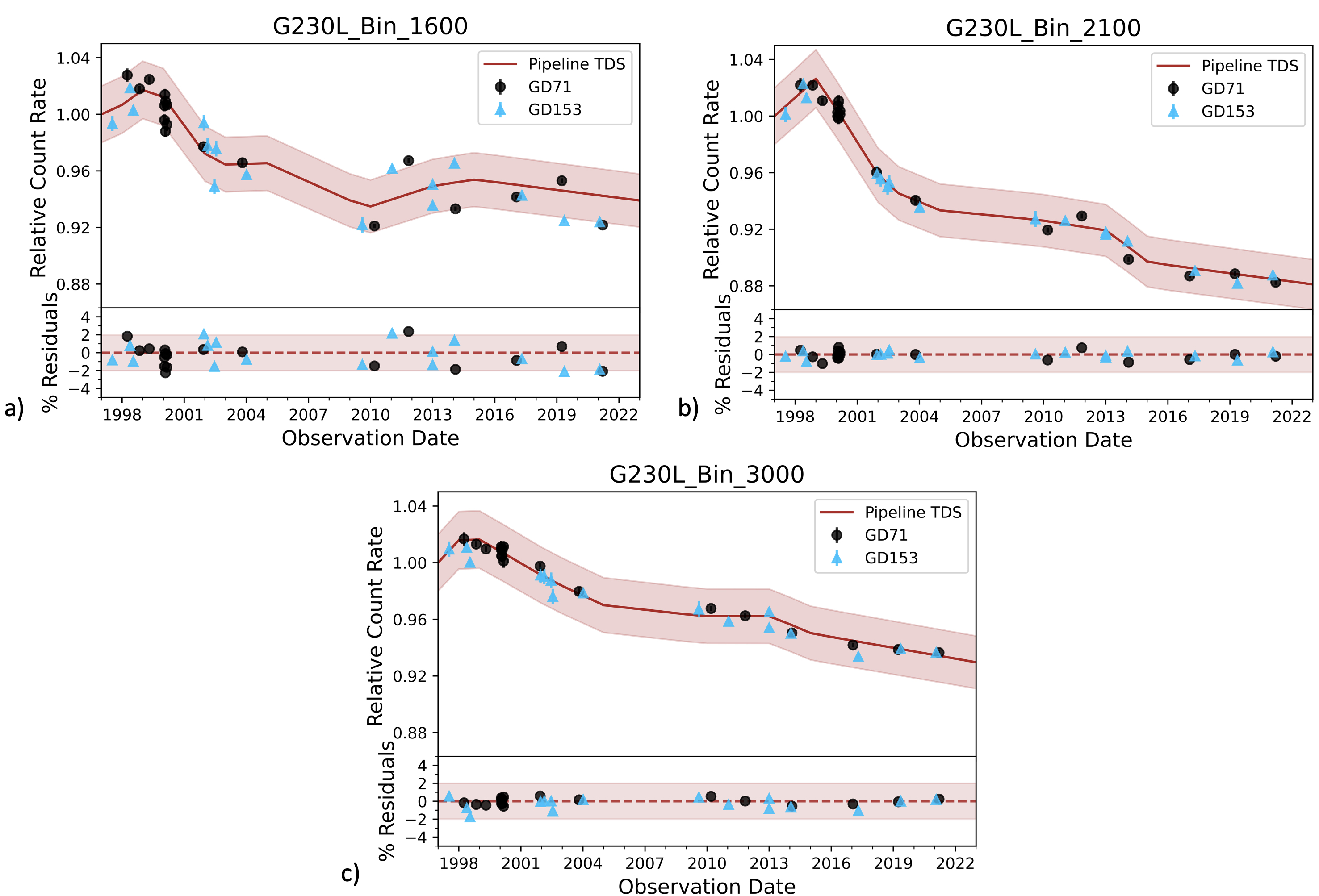}
    \caption{The same caption as \Cref{fig: Figure 7}, but for panels corresponding with G230L bins of starting wavelength: a) 1600 $\mathrm{\AA}$, b) 2100 $\mathrm{\AA}$, and c) 3000 $\mathrm{\AA}$, respectively.}
    \label{fig: Figure 8}
\end{figure}

\begin{figure}[!htbp]
  \centering
  \includegraphics[width=\textwidth]{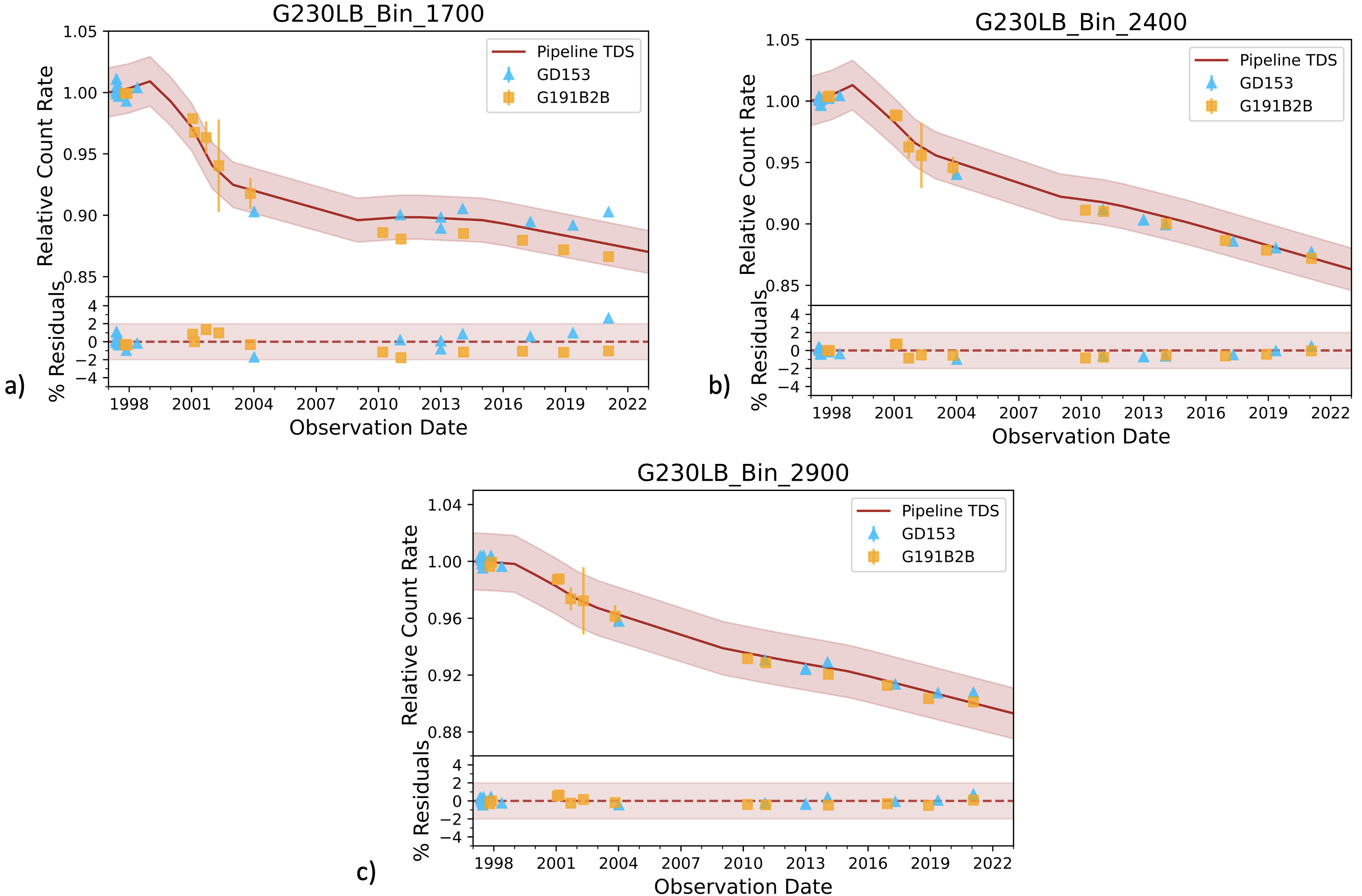}
    \caption{The same caption as \Cref{fig: Figure 7}, but for panels corresponding with G230LB bins of starting wavelength: a) 1700 $\mathrm{\AA}$, b) 2400 $\mathrm{\AA}$, and c) 2900 $\mathrm{\AA}$, respectively, and triad stars GD153 and G191B2B.}
    \label{fig: Figure 9}
\end{figure}

\begin{figure}[!htbp]
  \centering
  \includegraphics[width=\textwidth]{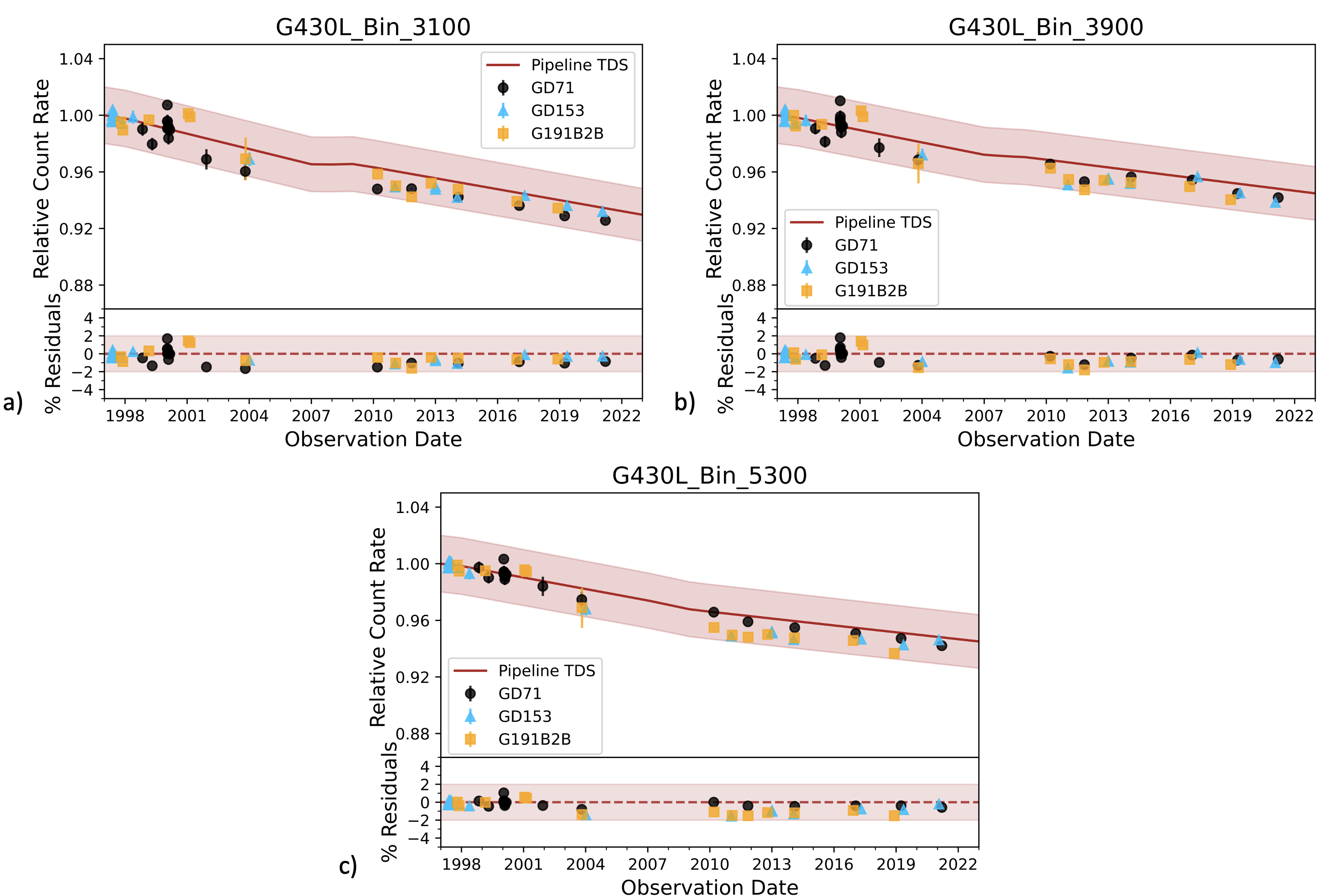}
    \caption{The same caption as \Cref{fig: Figure 7}, but for panels corresponding with G430L bins of starting wavelength: a) 3100 $\mathrm{\AA}$, b) 3900 $\mathrm{\AA}$, and c) 5300 $\mathrm{\AA}$, respectively, and all three triad stars.}
    \label{fig: Figure 10}
\end{figure}

\begin{figure}[!htbp]
  \centering
  \includegraphics[width=\textwidth]{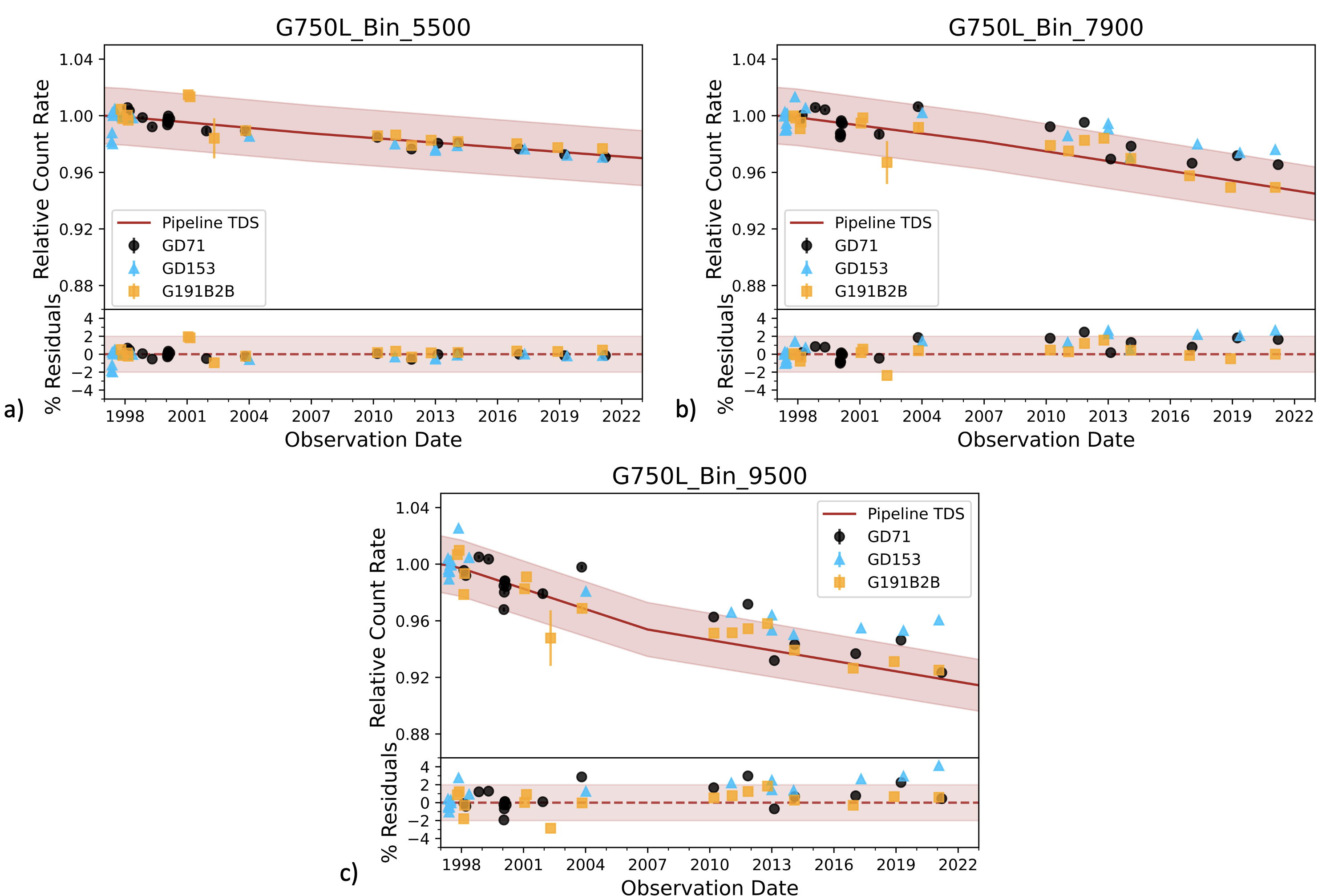}
    \caption{The same caption as \Cref{fig: Figure 7}, but for panels corresponding with G750L bins of starting wavelength: a) 5500 $\mathrm{\AA}$, b) 7900 $\mathrm{\AA}$, and c) 9500 $\mathrm{\AA}$, respectively, and all three triad stars. Along with \Cref{fig: Figure 6}, note that there is additional scatter for Post-Servicing Mission 4 data at redder wavelengths. This is observed with the TDS monitoring stars as well (STIS ISRs 2017-06 and 2021-01).}
    \label{fig: Figure 11}
\end{figure}

\newpage
\vspace{-0.3cm}
\ssectionstar{Appendix B}\label{sec:Appendix_B}
\vspace{-0.3cm}
The following table contains the unweighted average and standard deviation ($\sigma$) of the residuals in Figures 2-11. The bin number in $\mathrm{\AA}$ indicates the starting wavelength of that bin. The residuals are between the relative sensitivity of the standard star triad and the TDS pipeline trends. The standard deviation of the residuals is a measure of the spread of the percent difference between the triad net count rates and the TDS pipeline trends throughout the lifetime of STIS.
Where no data are available, a ``--'' is included. The ``All Data" column contains the unweighted average residuals of all available exposures for a given grating and wavelength bin. An unweighted average is used due to the similarity in data sample size and standard deviation for each triad star.

\begin{deluxetable}{cccccc}
    \tabcolsep 4pt
    \tablewidth{0pt}
    \tablecaption{The average percentage agreement between the standard star triad (GD71, GD153, and G191B2B) data and the adopted STIS TDS pipeline trends. The average residual and standard deviation ($\sigma$) for a triad star's data in units of percentage are provided where available. The unweighted average residuals of all available data for a given grating and wavelength bin are also included. \label{tab: Table 2}}
    \tabletypesize{\footnotesize}
    \tablecolumns{6}
    \tablehead{
        & & \multicolumn{4}{c}{Average Residual $\pm$ $\sigma$ ($\%$)}\\ 
        \colhead{Grating} & \colhead{Bin ($\mathrm{\AA}$)} & \colhead{GD71} & \colhead{GD153} & \colhead{G191B2B} & \colhead{All Data}}
    \startdata
    G140L & 1150 & 0.23 $\pm$ 1.03 & $-$0.08 $\pm$ 1.18 & -- & 0.09 $\pm$ 1.11 \\
      & 1200 & $-$0.05 $\pm$ 0.78 & $-$0.01 $\pm$ 0.59 & -- & $-$0.03 $\pm$ 0.70 \\
      & 1250 & 0.06 $\pm$ 0.68 & 0.17 $\pm$ 0.55 & -- & 0.11 $\pm$ 0.62 \\
      & 1300 & 0.13 $\pm$ 0.86 & $-$0.01 $\pm$ 0.66 & -- & 0.07 $\pm$ 0.78 \\
      & 1350 & 0.03 $\pm$ 0.86 & 0.04 $\pm$ 0.68 & -- & 0.03 $\pm$ 0.78 \\
      & 1400 & 0.25 $\pm$ 0.92 & 0.35 $\pm$ 0.91 & -- & 0.30 $\pm$ 0.92 \\
      & 1450 & 0.15 $\pm$ 0.87 & 0.24 $\pm$ 0.89 & -- & 0.19 $\pm$ 0.88 \\
      & 1500 & 0.15 $\pm$ 0.99 & 0.03 $\pm$ 0.91 & -- & 0.10 $\pm$ 0.96 \\
      & 1550 & 0.12 $\pm$ 1.26 & 0.22 $\pm$ 1.26 & -- & 0.17 $\pm$ 1.26 \\
      & 1600 & 0.09 $\pm$ 0.98 & $-$0.08 $\pm$ 0.93 & -- & 0.01 $\pm$ 0.96 \\
      & 1650 & 0.32 $\pm$ 1.11 & $-$0.04 $\pm$ 1.16 & -- & 0.15 $\pm$ 1.15 \\
      \hline \vspace{-0.3cm} \\
    G230L & 1600 & $-$0.34 $\pm$ 1.26 & $-$0.18 $\pm$ 1.36 & -- & $-$0.26 $\pm$ 1.31 \\
      & 1700 & $-$0.27 $\pm$ 0.56 & $-$0.13 $\pm$ 0.53 & -- & $-$0.2 $\pm$ 0.55 \\
      & 1800 & 0.03 $\pm$ 0.63 & 0.09 $\pm$ 0.66 & -- & 0.06 $\pm$ 0.64 \\
      & 1900 & $-$0.14 $\pm$ 0.55 & $-$0.24 $\pm$ 0.68 & -- & $-$0.19 $\pm$ 0.62 \\
      & 2000 & $-$0.18 $\pm$ 0.54 & $-$0.39 $\pm$ 0.59 & -- & $-$0.28 $\pm$ 0.57 \\
      & 2100 & $-$0.10 $\pm$ 0.5 & $-$0.04 $\pm$ 0.35 & -- & $-$0.07 $\pm$ 0.44 \\
      & 2200 & 0.26 $\pm$ 0.54 & 0.18 $\pm$ 0.36 & -- & 0.22 $\pm$ 0.47 \\
      & 2300 & 0.05 $\pm$ 0.43 & $-$0.02 $\pm$ 0.39 & -- & 0.02 $\pm$ 0.41 \\
      & 2400 & $-$0.06 $\pm$ 0.32 & $-$0.03 $\pm$ 0.32 & -- & $-$0.04 $\pm$ 0.32 \\
      & 2500 & $-$0.16 $\pm$ 0.3 & $-$0.13 $\pm$ 0.43 & -- & $-$0.14 $\pm$ 0.37 \\
      & 2600 & $-$0.09 $\pm$ 0.33 & $-$0.18 $\pm$ 0.42 & -- & $-$0.13 $\pm$ 0.38 \\
      & 2700 & $-$0.15 $\pm$ 0.34 & $-$0.22 $\pm$ 0.46 & -- & $-$0.18 $\pm$ 0.40 \\
      & 2800 & $-$0.08 $\pm$ 0.45 & 0.14 $\pm$ 0.49 & -- & 0.02 $\pm$ 0.49 \\
      & 2900 & $-$0.07 $\pm$ 0.33 & 0.15 $\pm$ 0.47 & -- & 0.04 $\pm$ 0.42 \\
      & 3000 & $-$0.01 $\pm$ 0.36 & $-$0.29 $\pm$ 0.63 & -- & $-$0.14 $\pm$ 0.52 \\
      \hline \vspace{-0.3 cm} \\
    G230LB & 1700 & -- & 0.24 $\pm$ 0.9 & $-$0.39 $\pm$ 0.93 & $-$0.02 $\pm$ 0.96 \\
      & 1800 & -- & $-$0.10 $\pm$ 0.85 & $-$0.70 $\pm$ 0.61 & $-$0.35 $\pm$ 0.81 \\
      & 1900 & -- & $-$0.40 $\pm$ 0.71 & $-$0.43 $\pm$ 0.58 & $-$0.41 $\pm$ 0.66 \\
      & 2000 & -- & $-$0.56 $\pm$ 0.78 & $-$0.51 $\pm$ 0.60 & $-$0.54 $\pm$ 0.72 \\
      & 2100 & -- & $-$0.31 $\pm$ 0.65 & $-$0.26 $\pm$ 0.62 & $-$0.29 $\pm$ 0.64 \\
      & 2200 & -- & $-$0.31 $\pm$ 0.57 & $-$0.26 $\pm$ 0.55 & $-$0.29 $\pm$ 0.56 \\
      & 2300 & -- & $-$0.35 $\pm$ 0.56 & $-$0.22 $\pm$ 0.50 & $-$0.30 $\pm$ 0.54 \\
      & 2400 & -- & $-$0.20 $\pm$ 0.42 & $-$0.28 $\pm$ 0.50 & $-$0.23 $\pm$ 0.45 \\
      & 2500 & -- & $-$0.20 $\pm$ 0.54 & $-$0.33 $\pm$ 0.47 & $-$0.25 $\pm$ 0.51 \\
      & 2600 & -- & $-$0.15 $\pm$ 0.39 & $-$0.12 $\pm$ 0.39 & $-$0.14 $\pm$ 0.39 \\
      & 2700 & -- & $-$0.19 $\pm$ 0.43 & $-$0.08 $\pm$ 0.40 & $-$0.14 $\pm$ 0.42 \\
      & 2800 & -- & $-$0.18 $\pm$ 0.42 & $-$0.11 $\pm$ 0.36 & $-$0.15 $\pm$ 0.40 \\
      & 2900 & -- & 0.02 $\pm$ 0.34 & $-$0.11 $\pm$ 0.36 & $-$0.03 $\pm$ 0.36 \\
      \hline \vspace{-0.3 cm} \\
    G430L & 2900 & 0.07 $\pm$ 0.97 & 0.19 $\pm$ 0.84 & 0.21 $\pm$ 0.79 & 0.15 $\pm$ 0.88 \\
      & 3100 & $-$0.50 $\pm$ 0.86 & $-$0.29 $\pm$ 0.42 & $-$0.31 $\pm$ 0.83 & $-$0.37 $\pm$ 0.72 \\
      & 3300 & $-$0.36 $\pm$ 0.83 & $-$0.14 $\pm$ 0.37 & $-$0.33 $\pm$ 0.75 & $-$0.27 $\pm$ 0.67 \\
      & 3500 & $-$0.35 $\pm$ 0.81 & $-$0.53 $\pm$ 0.60 & $-$0.48 $\pm$ 0.80 & $-$0.46 $\pm$ 0.74 \\
      & 3700 & $-$0.43 $\pm$ 0.83 & $-$0.27 $\pm$ 0.65 & $-$0.33 $\pm$ 0.80 & $-$0.34 $\pm$ 0.76 \\
      & 3900 & $-$0.25 $\pm$ 0.76 & $-$0.37 $\pm$ 0.53 & $-$0.54 $\pm$ 0.89 & $-$0.37 $\pm$ 0.73 \\
      & 4100 & $-$0.31 $\pm$ 0.74 & $-$0.49 $\pm$ 0.61 & $-$0.50 $\pm$ 0.88 & $-$0.43 $\pm$ 0.74 \\
      & 4300 & $-$0.28 $\pm$ 0.66 & $-$0.57 $\pm$ 0.65 & $-$0.62 $\pm$ 0.90 & $-$0.48 $\pm$ 0.74 \\
      & 4500 & $-$0.20 $\pm$ 0.60 & $-$0.30 $\pm$ 0.45 & $-$0.55 $\pm$ 0.84 & $-$0.33 $\pm$ 0.64 \\
      & 4700 & $-$0.25 $\pm$ 0.65 & $-$0.45 $\pm$ 0.55 & $-$0.59 $\pm$ 0.91 & $-$0.41 $\pm$ 0.71 \\
      & 4900 & $-$0.26 $\pm$ 0.59 & $-$0.48 $\pm$ 0.65 & $-$0.58 $\pm$ 0.83 & $-$0.43 $\pm$ 0.69 \\
      & 5100 & $-$0.13 $\pm$ 0.55 & $-$0.47 $\pm$ 0.67 & $-$0.58 $\pm$ 0.77 & $-$0.38 $\pm$ 0.68 \\
      & 5300 & $-$0.16 $\pm$ 0.40 & $-$0.45 $\pm$ 0.56 & $-$0.73 $\pm$ 0.73 & $-$0.42 $\pm$ 0.60 \\
      & 5500 & $-$0.10 $\pm$ 0.44 & $-$0.71 $\pm$ 0.78 & $-$0.51 $\pm$ 0.68 & $-$0.44 $\pm$ 0.70 \\
      \hline \vspace{-0.3 cm} \\
    G750L & 5500 & $-$0.03 $\pm$ 0.31 & $-$0.31 $\pm$ 0.67 & 0.29 $\pm$ 0.70 & $-$0.04 $\pm$ 0.63 \\
      & 5900 & 0.11 $\pm$ 0.30 & $-$0.10 $\pm$ 0.45 & 0.19 $\pm$ 0.66 & 0.06 $\pm$ 0.50 \\
      & 6300 & 0.03 $\pm$ 0.28 & $-$0.02 $\pm$ 0.34 & 0.13 $\pm$ 0.60 & 0.04 $\pm$ 0.42 \\
      & 6700 & $-$0.04 $\pm$ 0.32 & 0.04 $\pm$ 0.35 & 0.04 $\pm$ 0.64 & 0.01 $\pm$ 0.45 \\
      & 7100 & $-$0.11 $\pm$ 0.42 & 0.08 $\pm$ 0.44 & 0.05 $\pm$ 0.53 & 0.00 $\pm$ 0.47 \\
      & 7500 & $-$0.01 $\pm$ 0.52 & 0.05 $\pm$ 0.59 & 0.02 $\pm$ 0.61 & 0.02 $\pm$ 0.58 \\
      & 7900 & 0.57 $\pm$ 1.00 & 0.70 $\pm$ 1.26 & 0.07 $\pm$ 0.85 & 0.46 $\pm$ 1.09 \\
      & 8300 & 0.52 $\pm$ 0.97 & 0.51 $\pm$ 1.13 & $-$0.04 $\pm$ 0.86 & 0.35 $\pm$ 1.03 \\
      & 8700 & 0.51 $\pm$ 1.11 & 0.69 $\pm$ 1.32 & $-$0.11 $\pm$ 0.97 & 0.39 $\pm$ 1.20 \\
      & 9100 & 0.56 $\pm$ 1.19 & 0.53 $\pm$ 0.91 & 0.07 $\pm$ 1.03 & 0.40 $\pm$ 1.07 \\
      & 9500 & 0.54 $\pm$ 1.23 & 1.07 $\pm$ 1.42 & 0.23 $\pm$ 1.13 & 0.64 $\pm$ 1.32 \\
      \hline \vspace{-0.3 cm} \\
    \enddata
\end{deluxetable}

\end{document}